\documentclass[twocolumn,pra]{revtex4-1}

 \usepackage{amssymb}
 \usepackage{dsfont}
 \usepackage{mathdots}
 \usepackage{amsmath}
\usepackage{graphicx}
\usepackage{dcolumn}
\usepackage{bm}
\usepackage{color}

\begin{document}

\title{Coherent control of multipartite entanglement}
\author{Seyed~Mohammad~Hashemi~Rafsanjani, and Joseph H. Eberly}
\affiliation{Center for Coherence and Quantum Optics and the Department of Physics \& Astronomy, University of Rochester, Rochester, New York 14627, USA}
\email{hashemi@pas.rochester.edu}


\date{\today}

\begin{abstract} 
Quantum entanglement between an arbitrary number of remote qubits is examined analytically. 
We show that there is a non-probabilistic way to address in one context the management of entanglement of an arbitrary number of mixed-state qubits by engaging quantitative measures of entanglement and a specific external control mechanism. Both all-party entanglement and weak inseparability are considered. We show that for $N\ge 4$, the death of all-party entanglement is permanent after an initial collapse. In contrast, weak inseparability can be deterministically managed for an arbitrarily large number of qubits almost indefinitely. Our result suggests a picture of the path that the system traverses in the Hilbert space. 
\end{abstract}


\maketitle

 \newcommand{\beq}{\begin{equation}}
 \newcommand{\eeq}{\end{equation}}
 \newcommand{\bel}{\begin{align*}}
 \newcommand{\tamam}{\end{align*}}
 \newcommand{\dg}[1]{#1^{\dagger}}
 \newcommand{\reci}[1]{\frac{1}{#1}}
 \newcommand{\ket}[1]{|#1\rangle}
 \newcommand{\nim}{\frac{1}{2}}
 \newcommand{\om}{\omega}
 \newcommand{\te}{\theta}
 \newcommand{\la}{\lambda}
 \newcommand{\beqa}{\begin{eqnarray}}             
 \newcommand{\eeqa}{\end{eqnarray}}               
 \newcommand{\nn}{\nonumber}                      
 \newcommand{\bra}[1]{\langle#1\vert}                 
 \newcommand{\ipr}[2]{\left\langle#1|#2\right\rangle}
 \newcommand{\up}{\uparrow}
 \newcommand{\down}{\downarrow}
  \newcommand{\dn}{\downarrow}         

\section{introduction}
Despite recent advances in experimental realization of multipartite entanglement \cite{PhysRevLett.106.130506}, current schemes to preserve entanglement, such as the quantum Zeno effect \cite{PhysRevLett.100.090503}, entanglement distillation \cite{PhysRevA.53.2046, *Kwiat:2001bv, *Pan:2003kv,*Dong:2008cj} or weak measurements \cite{Natureweak}, lack an element of control and/or their success is probabilistic. We show here  that there is a way to address all five aspects of the managed entanglement question, namely obtaining in one prescription simultaneous compatibility of (i) mixed states, (ii) arbitrary numbers of qubits, (iii) quantitative measure of entanglement, (iv) non-probabilitic success, and (v) external control. We show that the phenomenon of collapse and revival \cite{eberly1980} offers a concrete example of a mechanism of deterministic control of multi-qubit mixed-state entanglement.  


Here we combine knowledge about $N$-party entanglement with control of  revival dynamics to demonstrate quantitative control of multipartite entanglement \cite{[{Elements of our procedure have only been exploited for two-qubits previously: }] yonac2010}.
We present an example of multipartite entanglement that is 
initially shared by $N$ remote qubits interacting with individual 
fields. Local control is managed by a coherent state of a resonant 
mode via collapse and revivals of the qubit coherences.
By controlling the amplitude of the coherent states one controls the time of revivals and thus the recovery of the multi-qubit entanglement. 
We note that entanglement and revivals were previously discussed in studies focused on the entanglement of one qubit and its local field \cite{PhysRevLett.65.3385,PhysRevA.44.6023}. Here, we mean the entanglement among the qubits and not inseparability from their local fields. 

Multipartite entanglement can signal inseparability for different partitionings of the system. Here we examine two extreme kinds of multipartite entanglement: (i) all-party entanglement, also known as genuinely multipartite entanglement, which signals inseparability along all possible partitionings, and (ii) weak inseparability, defined as the lack of full separability. Full separability signals that the state is not entangled along any partitioning. We develop an approximation that allows us to obtain analytical expressions for both of these quantities. 
For a quantitative analysis we need to quantify the two kinds of entanglement: all-party entanglement and weak inseparability. For all-party entanglement, there have been advances in determining whether a state is entangled or not \cite{Guhne2009,1367-2630-12-5-053002,PhysRevLett.104.210501,PhysRevA.83.062325,PhysRevA.83.062337,PhysRevLett.108.230502,PhysRevLett.108.020502,Eltschka:2014wm}. We avoid numerical approaches since the dimension of Hilbert space grows prohibitively large  \cite{PhysRevLett.106.190502,*PhysRevA.88.012305}. We pay particular attention to the special case of $N$-qubit X-states  \cite{[{Properties of X-states were first exploited in the two-qubit context:}] Xmatrix}. \citet{PhysRevA.86.062303} have developed an algebraic formula for their all-party entanglement. Remarkably the entanglement of the X-part of any $N$-qubit density matrix is a lower bound for the entanglement of the complete matrix \cite{PhysRevA.83.062325,Rafsanjani-arxiv}.

The qubits are initially assumed to be in a Greenberger-Horne-Zeilinger (GHZ) state \cite{PhysRevLett.82.1345} and we explain below an approximation that reduces their density matrix to an X-state for all times. X-states are $N$-qubit density matrices whose non-zero elements are restricted to diagonal or anti-diagonal in an orthonormal product basis. They include important states such as GHZ and GHZ-diagonal states. Our approximation enables us to  use the algebraic formula developed in \cite{PhysRevA.86.062303} to quantify the all-party entanglement. We also utilize the distance from the set of fully-separable states as our measure of weak inseparability. We obtain an analytical formula for this quantity during dynamics. We observe that beyond three qubits the initial loss of all-party entanglement after collapse is permanent. We then examine weak inseparability, and demonstrate that, contrary to all-party entanglement, weak inseparability experiences revivals even for very large values of $N$, although the strength of such revivals decreases with $N$. Our result suggests a clear picture of the path that the $N$-qubit state traverses during the dynamics. Last but not least we make an attempt to capture the distribution of the entanglement during the collapse interval. It is shown that at the middle of a collapse interval the initial entanglement is completely transferred to entanglement between resonators. 

Each local field is described by a resonant mode of the field that is interacting with its local qubit. To observe revivals one has to prepare a resonator with a very small leakage constant. In our case this means $N$ times smaller than the leakage constant needed to observe a revival in a single resonator. For a coherent state $\ket{\alpha}$ with $\alpha^{2}=100$ a ratio $10^{3}$ of coupling constant to the decay rate is required, which is not outrageously higher than $3\times10^{2}$, that was achieved in a circuit QED setup recently \cite{Vlastakis01112013}. Thus only an order of magnitude improvement in this ratio leads to suitable condition for control of revival of multipartite entanglement in a setup with $N\ge 3$. We assume this condition is satisfied and ignore the resonator leakage altogether. 

\section{Collapse and revival in Jaynes-Cummings model}
Each of $N$ remote and identical subsystems is made of a two-level system (a qubit) that is interacting with a single mode of the electromagnetic field through a Jaynes-Cummings interaction \cite{jcpaper}.  
\begin{align}
H_i=\frac{\omega_{0}}{2}\sigma_{zi}+g(a_i^{\dagger}\sigma^-_{i}+a_i\sigma^+_{i})+\omega a_i^{\dagger}a_i
\end{align}
The JC Hamiltonian is integrable and one can analytically follow the evolution of the above model. In the following, and for simplicity, we further assume that the qubit is in resonance with the resonator, i.e. $\omega_0=\omega$.

A coherent state can be written 
as $\ket{\alpha}=\sum_n A_n \ket{n}$ and $A_n=\exp(-\alpha^2/2)\alpha^n/\sqrt{n!}$ where $\ket{n}$ is a Fock state of $n$ excitations. For simplicity we assume that $\alpha$ is real and positive. Then we can write the coherently driven evolution of $\ket{e}$ and $\ket{g}$ states:
\begin{align}
U_t\ket{e,\alpha}&=\ket{e,\alpha}_t=\ket{e} \otimes  \ket{\phi_0}+\ket{g} \otimes \ket{\phi_1}\\ \nn
U_t\ket{g,\alpha}&=\ket{g,\alpha}_t=\ket{e} \otimes  \ket{\phi_2}+\ket{g} \otimes \ket{\phi_3}
\end{align}
where $\ket{\phi_{i}}=\sum_n \phi_{in}\ket{n}$ and their coefficients are $\phi_{0n}=A_{n}r_{n+1},~\phi_{1n}=A_{n-1}t_{n},~\phi_{2n}=A_{n+1}t_{n+1},~\phi_{3n}=A_{n}r_{n}$, $r_{n}=e^{-i\omega t}\cos(gt\sqrt{n})$, and $t_{n}=-ie^{-i\omega t}\sin(gt\sqrt{n})$. All the dynamics we intend to investigate can be captured by the inner products of these four different $\ket{\phi_i}$'s. For $\alpha\ge10$ excellent approximations are available to evaluate these inner products. The summary of the results is given below.
 \begin{align}\nn
  &\ipr{\phi_i}{\phi_i}=\frac{1+p_1x}{2},~~~~~~~~~~~~~~~p_1=\big\{_{-1~~~~i=1,2}^{+1~~~~i=0,3} \\ \nn
  &\ipr{\phi_i}{\phi_{i+1}}=i \frac{e^{i\omega t}}{2}(I_{2}+p_2y),~~~p_2=\big\{_{-1~~~~i=0}^{+1~~~~i=2} \\ \nn
  &\ipr{\phi_i}{\phi_{3-i}}= \frac{e^{i\omega t}}{2} (I_{1}+p_3x),~~~~p_3=\big\{_{-1~~~~i=0}^{+1~~~~i=2} \\ \nn
&\ipr{\phi_0}{\phi_2}=-\ipr{\phi_1}{\phi_3}=-\frac{iy}{2} 
\end{align}
where $x+iy=\sum_n A_n^2\exp(2igt\sqrt{n})$, and 
\begin{align}
I_{1}+iI_{2}\simeq\exp(-\frac{g^{2}t^{2}}{32\alpha^{4}}) e^{\frac{igt}{2\alpha}}.
\end{align}

These four quantities are evaluated in Appendix A. In Fig.~\ref{xplot} we present the time dependence of $x, I_1$, and $I_2$ for the coherent states with $\alpha=10$. The plot for $y$ is similar to the plot for $x$ except that the fast Rabi oscillations in the revivals are $\pi$ out of phase. At the first revival the maximum of $|x|$ is $\frac{1}{2}$, and $I_1\simeq -1$. At all revivals $I_2\simeq 0$.

\begin{figure}[htbp]
\begin{center}
\includegraphics[width=\columnwidth]{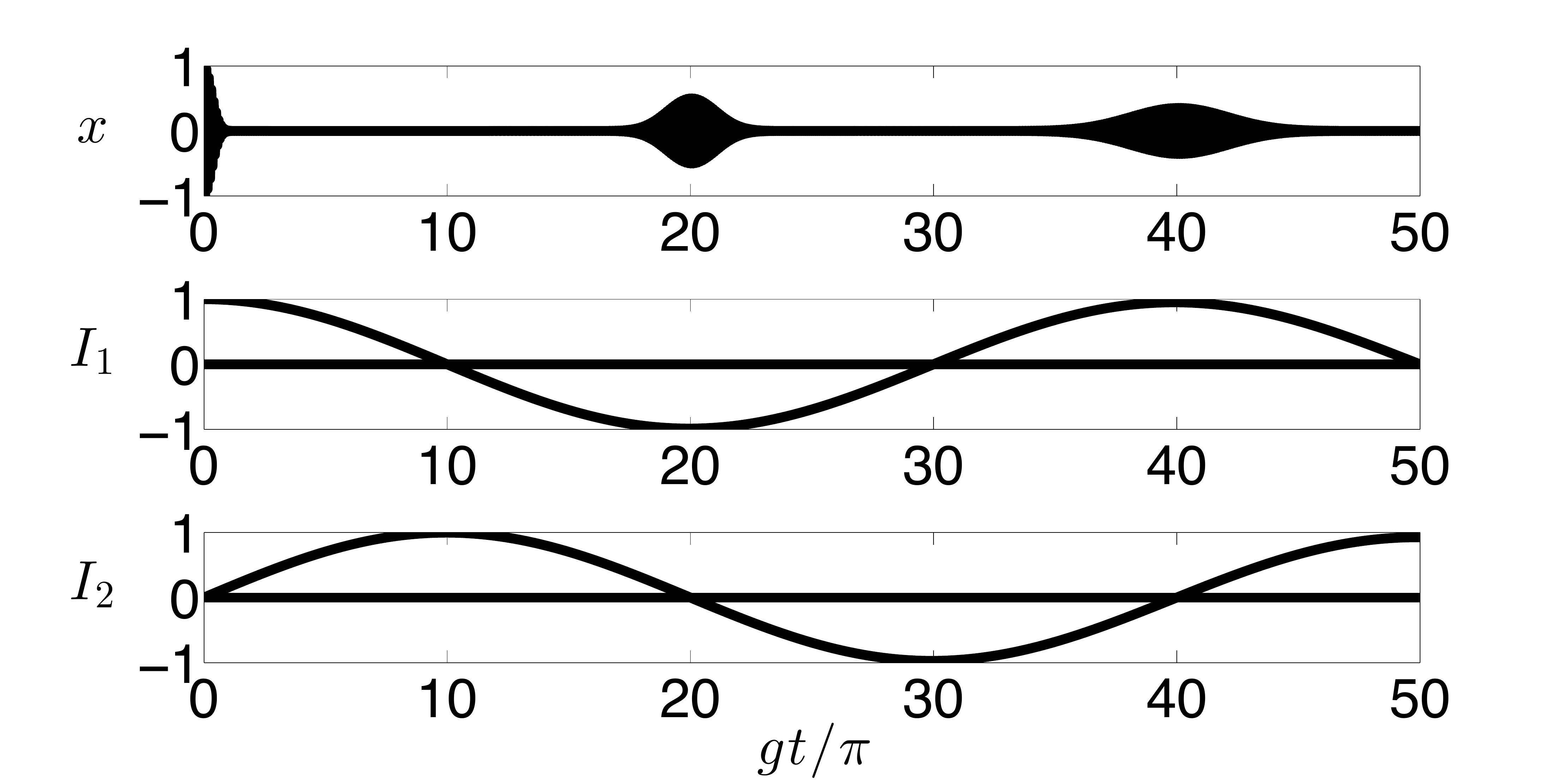}
\vspace{-0.75cm}
\caption{The time dependence of $x, I_1$, and $I_2$ for coherent states of $\alpha=10$.}
\label{xplot}
\end{center}
\vspace{-0.5cm}
\end{figure}

\section{Density matrices}
In this section we derive the density matrix of the qubits. The qubits are initially in a GHZ state and the resonators are in identical coherent states:
\begin{align}
\left(\frac{\ket{e,e,\cdots,e}+\ket{g,g,\cdots,g}}{\sqrt{2}}\right)\otimes \ket{\alpha,\alpha, \cdots, \alpha}.
\end{align}
We introduce a simplifying approximation that allows for omission of many elements of the density matrix. The approximation guarantees that the $N$-qubit density matrix will  remain  an X-state. X-states are $N$-qubit states whose non-zero 
elements are restricted to diagonal or anti-diagonal:
\begin{align}
\hat{X}=\left(  \begin{array}{cccccccc}
    a_{1} & ¥ & ¥ & ¥ & ¥ & ¥ & ¥ & z_{1} \\ 
    ¥ & a_{2} & ¥  & ¥ & ¥ & ¥ & z_{2} & ¥ \\ 
    ¥ & ¥ & \ddots & ¥ & ¥ & \iddots & ¥ & ¥ \\ 
    ¥ & ¥ & ¥ & a_{d} & z_{d} & ¥ & ¥ & ¥ \\ 
    ¥ & ¥ & ¥ & z_{d}^{*} & b_{d} & ¥ & ¥ & ¥ \\ 
    ¥ & ¥ & \iddots & ¥ & ¥ & \ddots & ¥ & ¥ \\ 
    ¥ & z_{2}^{*} & ¥ & ¥ & ¥ & ¥ & b_{2} & ¥ \\ 
    z_{1}^{*} & ¥ & ¥ & ¥ & ¥ & ¥ & ¥ & b_{1} \\ 
  \end{array}
\right).
\end{align}
Here~$d=2^{N-1}$, and we require $|z_{i}|\le \sqrt{a_{i}b_{i}}$ and $\sum_{i}(a_{i}+b_{i})=1$ to ensure that $\hat{X}$ is positive and normalized. We also denote $w_{i}=\sum_{j\neq i}^{d}\sqrt{a_{j} b_{j}}$. It is shown in \cite{PhysRevA.86.062303} that the all-party concurrence of an $N$-qubit X-matrix is
 \begin{align}
C_{N}= 2\max\{0,|z_{i}|-w_{i}\}, ~ i=0,1,\dots,d.
\end{align}
In deriving the density matrix, we follow the same approximation that leads to an X-state, and was developed by \citet{yonac2010}.
To decide which elements of the matrix can be safely discarded, we replace the coherent state $\ket{\alpha}$ by a Fock state $\ket{\bar{n}}$ where $\bar{n}=\alpha^2$, which is supported by the fact that for $\bar{n}\gg1$ the photon distribution of a coherent state is relatively narrowly peaked around $\bar{n}$. 
One can instead assume a less extreme variant of this approximation where $\ket{\alpha}$ is replaced by a mixture of Fock states $\ket{n}$ that has the same photon distribution as $\ket{\alpha}$. 
Both of these approximations lead to an important consequence: the density matrix of the qubits becomes an X-state with only one non-zero off-diagonal element.  

We then calculate the values of the non-zero elements using their values from a coherent state. 
In fact the simplification goes even further and we will show that $a_{i}=b_{i}$, i.e., the density matrix remains a GHZ-diagonal matrix.
The above approximation has been shown to be an excellent choice in capturing the collapse and revivals \cite{Yonac:08,yonac2010}.
Finally we emphasize that the estimates of entanglement that we derive from the above approximation are lower bounds of the entanglement of the complete matrix, where no elements were discarded \cite{PhysRevA.86.062303}.

To simplify the calculations we choose the initial entangled state of the qubits to be a symmetric state with respect to the permutation of the qubits. Next we calculate the non-zero elements of the density matrix. 
We use the orthonormal product basis $\{\ket{ee\dots e}, \ket{ee\dots g},\dots, \ket{gg\dots g}\}$ to represent the density matrices of the $N$ qubits and denote
$
\ket{e^{\otimes p}, g^{\otimes q}}=\otimes^p\ket{e}\otimes^q\ket{g}.$

We first focus on the case of two qubits \cite{yonac2010}. At any time the state of the system will read
\begin{align}\nn
\ket{\Psi}_t=\frac{1}{\sqrt{2}}\bigl( \otimes^2\ket{e,\alpha}_t+\otimes^2\ket{g,\alpha}_t \bigr).
\end{align}
The elements that we are interested in are 
\begin{align} \nn
 |X_{ee,gg}|=&\frac{1}{4}|I_1^2+x^2-y^2-I_2^2|, \\ \nn
|X_{eg,eg}|=& \frac{1}{4}(1-x^2+y^2).
\end{align}
We note that $X_{eg,eg}=X_{ge,ge}$. Now we discuss the case of $N\ge3$. The state of the system is given by 
\begin{align}
\ket{\Psi}_t=\frac{1}{\sqrt{2}}\bigl( \otimes^N\ket{e,\alpha}_t+\otimes^N\ket{g,\alpha}_t \bigr).
\end{align}
According to our approximation we only need to calculate one off-diagonal element:
\begin{align}\nn
2^{N+1}|X_{ee\cdots e, gg\cdots g}|&= \left|  (-i)^N((I_2-y)^N+(I_2+y)^N)+K \right|,\\
K&= (I_1-x)^N+(I_1+x)^N.
\end{align}
Next we calculate the diagonal elements:
\begin{align} \nn
&2X_{e^{\otimes n} g^{\otimes m},e^{\otimes n} g^{\otimes m}}=\frac{1}{2^{n+m}}[(iy)^{n+m}((-1)^n+(-1)^m) ]\\ 
&+\frac{1}{2^{n+m}} [    (1+x)^n(1-x)^m + (1-x)^n(1+x)^m].
\end{align}
This equation implies that $\bra{e^{\otimes p}, g^{\otimes q}} \hat{X} \ket{e^{\otimes p}, g^{\otimes q}} =\bra{g^{\otimes p}, e^{\otimes q}} \hat{X} \ket{g^{\otimes p}, e^{\otimes q}} .$
Using the above equations and also the permutation symmetry of the problem we can find all the diagonal elements of the density matrix. This simplification confirms that the X-part of the state will always remain a GHZ-diagonal state. This has two consequences. First the concurrence of a GHZ-diagonal state is directly proportional to the distance of that state to the set of biseparable states \cite{PhysRevA.88.062331}. This enables us to draw a picture of the trajectory that the state traverses in the Hilbert space.~Second, since these GHZ-diagonal states have only one non-zero anti-diagonal element, we can determine the full-separability of them \cite{PhysRevA.61.042314}.

\section{Entanglement Revival dynamics}
So far we have discussed the dynamics of collapse and revivals in the Jaynes-Cummings model as well as the derivation of the evolution of the $N$-qubit density matrix under appropriate assumptions. Now we turn our attention to  the multipartite entanglement. One has to be careful that going from the bipartite case to the multiqubit case, the number of partitionings of the $N$ parties grows more than one and inseparability can occur for different kinds of partitionings. Thus one first needs to clarify the kind of multipartite entanglement that is discussed.
 The hierarchy of multipartite entanglement that we are dealing with here comes from the concept of $k$-separability \cite{RevModPhys.81.865}. A pure state of $N$ parties (here qubits) is called $k$-separable if there is a $k$-partitioning along which the state is separable. That is to say that the pure state can be written as
\begin{align}
\ket{\psi}=\ket{\psi_{1}}\otimes\ket{\psi_{2}}\otimes\cdots \otimes \ket{\psi_{k}}
\end{align} 
To extend the definition to mixed states, a mixed state that can be written as a convex sum of $k$-separable pure states is defined to be $k$-separable. Note that by definition a $k+1$-separable state is also a $k$-separable state. The two extremes of this hierarchy are the following. In one end of the hierarchy are the states that are not even biseparable. Such states are referred to as possessing genuinely multipartite entanglement. In the current manuscript we refer to this entanglement as all-party entanglement. The measure that we use to quantify the all-party entanglement is the all-party concurrence that was first introduced by \cite{PhysRevA.83.062325}. Since our approximation has led to a GHZ-diagonal state,  we can take advantage of the formula we developed in \cite{PhysRevA.86.062303} to quantify the all-party concurrence.  We also note that in this case our measure has a geometrical interpretation that we will use in the following section. We have shown in \cite{PhysRevA.88.062331} that the all-party concurrence of GHZ-diagonal states is equal to the distance of the entangled state from the set of biseparable states where the distance is quantified by the trace distance \cite{Nielsen:2000}. 

In the other end of the hierarchy, there are states that are $N$-separable states, more generally referred to as fully-separable states. If all-party entanglement can be thought of as the most exclusive kind of entanglement whose presence implies that system is inseparable along any possible partitioning. Full separability is the other extreme, implying that the system is separable along all possible partitionings. An $N$-party system is fully separable if it can be written as $\sum_{i} p_{i}~~  \rho_{1i}\otimes \rho_{2i}\otimes \cdots \otimes \rho_{Ni}.$ The onset of full separability is the true end of entanglement. We call the lack of full-separability as weak inseparability. To quantify the weak inseparability we take advantage of the special form of the GHZ-diagonal states and use a geometrical measure that we developed in Appendix B. Our measure of weak inseparability is the distance from the set of fully-separable states, where the distance between two quantum states is given by their trace distance \cite{PhysRevA.88.062331}. In Appendix B we have derived an algebraic formula for the value of this measure for the GHZ-diagonal states that have only one non-zero off-diagonal element. This distance $S$ is a proper measure of weak inseparability and can be calculated for GHZ-diagonal states for which all but one of the anti-diagonal elements vanish:
\begin{align}
S=2\max\{0,|z_{1}|-c\},~~~~~c=\min\{b_{i}\}.
\label{inseparabilityformula}
\end{align}
Note that both of our measures to quantify the all-party entanglement and weak inseparability use the trace distance as the distance measure to the set of biseparable states and fully-separable states respectively.

\subsection{All-party entanglement}
\begin{figure}[tbp]
\begin{center}
\includegraphics[width=\columnwidth]{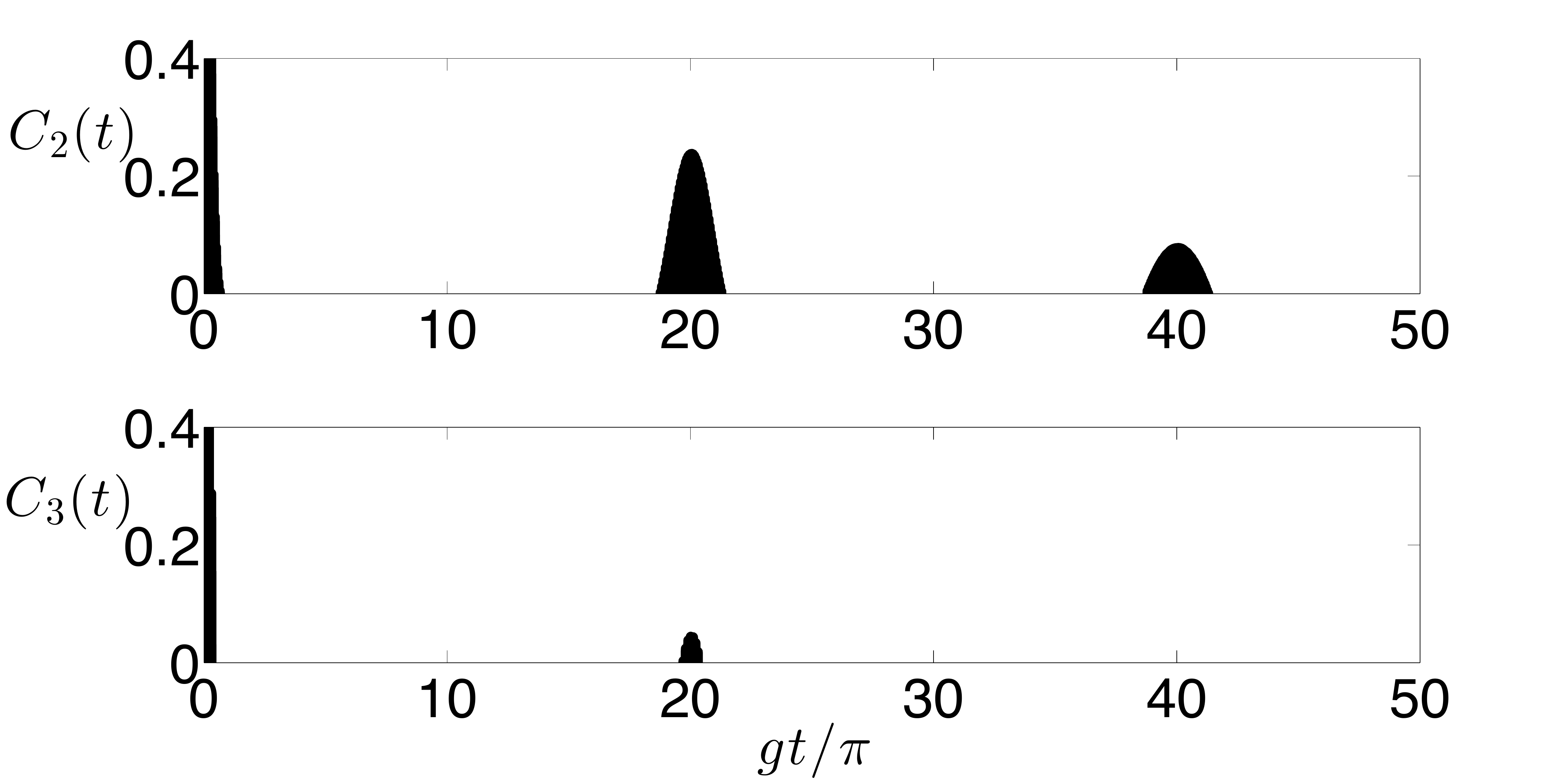}
\vspace{-0.5cm}
\caption{The revival of the entanglement for bipartite and genuinely tripartite entanglement for coherent states of $\alpha=10$.}
\label{qubit}
\end{center} \vspace{-0.5cm}
\end{figure}

For $N=2$ the result matches 
the result in \cite{yonac2010}, and the all-party concurrence is given by $C_{2}=\max\{0,Q_{2}\}$ where
\begin{align}
Q_{2}=\frac{1}{2}(I_1^2+2x^2-2y^2-I_2^2-1).\label{two-qubits}
\end{align}
For $N=3$ the all-party concurrence is $C_{3}=\max\{0,Q_{3}\}$ where $Q_{3}=2(|X_{eee,ggg}|-3|X_{eeg,eeg}|)$ and 
\begin{align} \nn
|X_{eee,ggg}|=&\frac{1}{8}\sqrt{(I_1^3+3I_1^2x)^2+(I_2^3+3I_2^2y)^2}\\
|X_{eeg,eeg}|=&\frac{1}{8}(1-x^2).
\end{align}
We plot $C_2$ and $C_3$ as a function of time in Fig.~\ref{qubit}.
As expected the entanglement dies out rapidly with the initial collapse. At $gt=2\pi \alpha$ entanglement revives to a small value.
The maximum of this revival can be estimated from the above equations noting that the maximum value of $x$ at first revival is $1/2$, and $I_1\simeq -1$. At the revivals $I_2\simeq0$ and since $x,y$ are completely out of phase with each other when $x$ is at maximum revival, $y$ vanishes. Feeding these quantities to the above equations leads to the maximum height of the bipartite and tripartite entanglement revivals to be $\frac{1}{4}$ and $\frac{1}{16}$ respectively. The second revival does not occur for tripartite entanglement. For $N>3$ the sudden death of all-party entanglement is permanent. 

\subsection{Weak inseparability}
\begin{figure}[t]
\begin{center}
\includegraphics[width=\columnwidth]{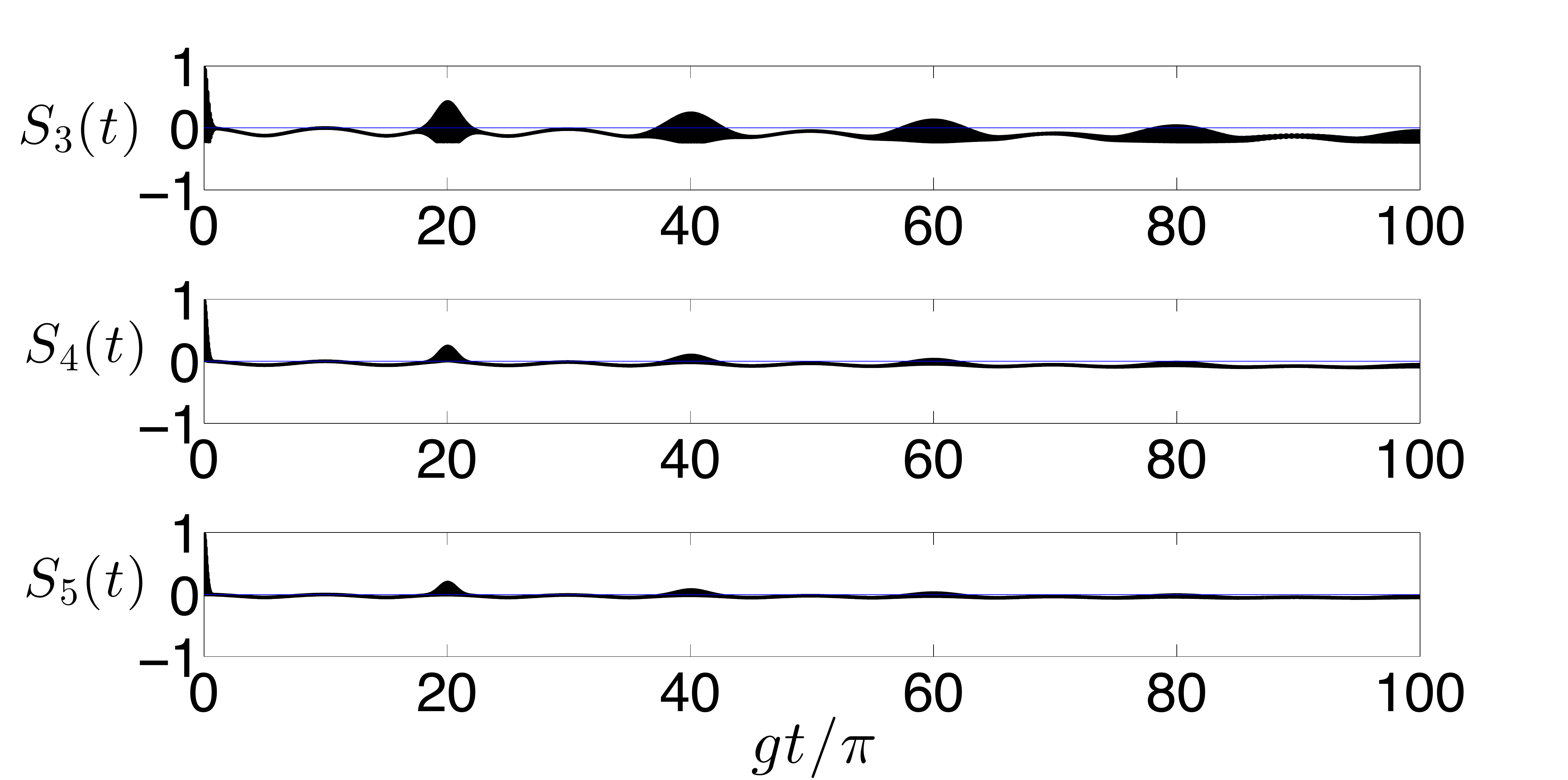}
\vspace{-0.55cm}
\caption{The revival of weak inseparability for $N=3,4,5$ for coherent states of $\alpha=10$. } 
\label{non-separability}
\end{center}
\vspace{-0.5cm}
\end{figure}

Now we turn our focus to weak inseparability.
We use the distance of the state from the set of fully separable states to study the dynamics of weak inseparability. 
In Fig.~\ref{non-separability} we plot the value of this distance for $N=3,4,5$. In contrast to the all-party entanglement  we see that weak inseparability revives at multiples of $gt=2\pi\alpha$ until it decays away completely. 
We can make the same observation even for large values of $N$. To this end we replace $c$ in the weak inseparability formula with the average values of all $b_{i}$'s. We also assume that $N$ is odd. Thus we can approximate 
\begin{align}
S_{N}\simeq \frac{1}{2^{N}}[(1-x)^{N}+(1+x)^{N}-2].
\end{align}
This has the strong implication that at these revivals this inseparability revives even for very large $N$. Yet its maximum distance from the boundary between fully separable and inseparable states decreases exponentially. For example at the first revival the maximum value of $x$ is $\frac{1}{2}$ and thus we can estimate $S_{N}\simeq (\frac{3}{4})^{N}$. Thus for large $N$ the state follows a path to the boundary of inseparability and then stays below it, crossing it momentarily only at revivals.

\begin{figure}[h]
\begin{center}
\includegraphics[width=7.5cm]{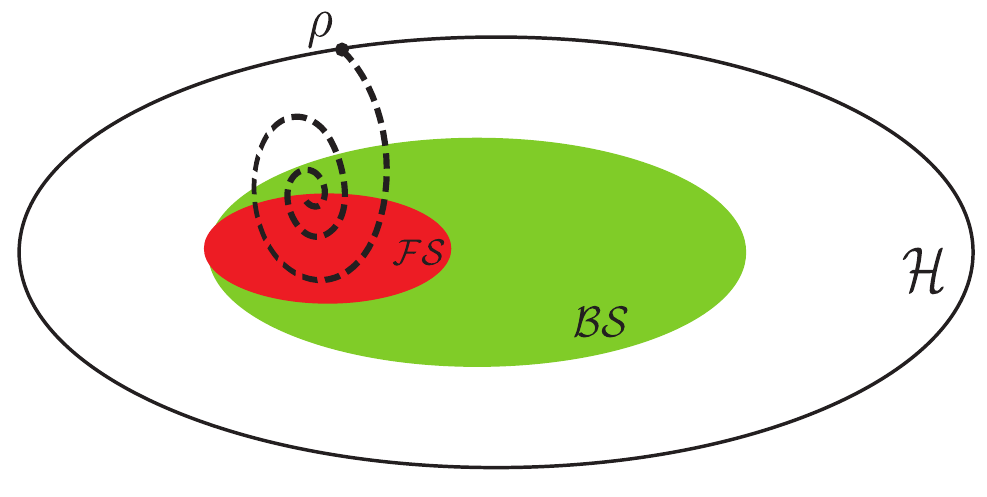}
\vspace{-0.25cm}
\caption{A schematic trajectory of the state in the Hilbert space for $N=3$. $\mathcal{H}$, $\mathcal{BS}$, and $\mathcal{FS}$ 
stand for the Hilbert space, biseparable subspace and fully separable states respectively. }
\label{schematic_path}
\end{center} \vspace{-0.5cm}
\end{figure}

The previous results and the fact that both of our measures have a geometrical interpretation can be combined to picture how the state moves in the Hilbert space \cite{Yu2007}. In the Hilbert space there the convex set of biseparable states, $\mathcal{BS}$. Inside this set there is the convex set of fully-separable states $\mathcal{FS}$. For $N=2$ these two sets match, and the initial state starts outside the $\mathcal{BS}$ and after five crossings it ends up inside $\mathcal{BS}$. For $N=3$ the state starts outside $\mathcal{BS}$ and moves inside $\mathcal{BS}$ and also inside $\mathcal{FS}$. At the first revival the state goes outside of both these sets and then comes back inside and gets trapped inside $\mathcal{BS}$ permanently. The state then moves inside $\mathcal{BS}$, crossing $\mathcal{FS}$ several times until it ends up somewhere in $\mathcal{FS}$. For $N\ge4$ once inside, the state never leaves $\mathcal{BS}$. It follows a trajectory that crosses $\mathcal{FS}$ several times and ends up in $\mathcal{FS}$. Fig.~\ref{schematic_path} provides a sketch of these dynamics.

\section{the flow of multipartite entanglement}
So far we have studied the dynamics of multipartite entanglement between the $N$ qubits. Since the collection of $N$ subsystems constitute a closed system and we started with a pure state, entanglement cannot flow out of the system. In fact the initial multipartite entanglement between the $N$ subsystem (qubit+resonators) should remain constant
since the dynamics is only a local unitary transformation for each subsystem. The question that arise is then how this entanglement is distributed between different qubits and resonators, when the entanglement between the qubits decreases. Is it stored as multipartite entanglement between resonators or is it stored as multipartite entanglement between different collections of qubits and the resonators. 

To answer this question may at first seem infeasible since the resonators are not qubits, and our measures of entanglement can only treat multiqubit states. However, we can derive some conclusions about the system at the center of the collapse period. At the moment when $I_{2}=1$, $I_{1}=0$, and $x=y=0$, the qubits and the resonator states become separable. This property was first pointed out by Gea-Banacloche \cite{PhysRevA.44.5913}. At that moment we can treat the resonators as instantaneous qubits and find the entanglement between them. The state of the system at $gt=\pi \alpha$ reads
\begin{align}
\otimes^{N}\left(\frac{\ket{e}+ie^{i\omega t}\ket{g}}{\sqrt{2}}\right)\otimes\bigl( \frac{\otimes^N\ket{\tilde{\phi}_{0}}+\otimes^N\ket{\tilde{\phi}_{2}}}{\sqrt{2}} \bigr).
\end{align}
where $\ket{\tilde{\phi_{i}}}=\ket{\phi_{i}}/\sqrt{\ipr{\phi_{i}}{\phi_{i}}}$ are normalized. Thus in the middle of collapse intervals all the initial entanglement is stored between the $N$ resonators. This behavior is similar to the distribution of entanglement if all the resonators were initially in the vacuum state. It is however noticeable that the time scale where such simplification is valid relates to the time scale where $I_{2}\simeq 1$, and $|I_{1}|\simeq 0$ which is proportional to 
$(\pi\alpha/g)$ which is $\alpha$ times longer than the time scale one associates with the period of entanglement for vacuum resonators $\pi/g$ for large $N$. 

\section{Conclusion}

In summary, we have provided an answer to an open
question in application of the principles of quantum information, 
namely how to exert a form of deterministic control over a 
quantitative degree of entanglement shared among an unspecified 
number of mixed-state qubits. Our answer is admittedly not perfect, 
and probably beyond near-term laboratory realization, but it is a 
strong step forward because it shows by concrete example that there 
can be a prescription that at the same time addresses all five 
difficult aspects of the question: mixed states, arbitrary numbers, 
quantitative measure, deterministic success, and external control, for entanglement of 
qubits.

We have used the machinery of quantifiable measures of entanglement 
to controllably suppress and recover specified degrees of 
multipartite entanglement using the phenomenon of coherent-state 
revivals. Our results are limited because we  explicitly studied only 
two extreme kinds of multipartite entanglement, namely all-party 
entanglement and weak inseparability. All-party entanglement is so 
fragile that beyond three qubits our method fails. But entanglement 
in the form of weak inseparability undergoes completely different 
dynamics and is almost indefinitely controllable. Even for a very 
large value of $N$, weak inseparability repeatedly revives from zero to a 
substantial non-zero value before disappearing again, although the 
strength of the revivals shrinks with $N$. Our 
results suggest a picture of system evolution in the Hilbert space -- 
if the system starts in a genuinely $N$-partite entangled state its evolution takes it
back and forth over the boundary of full-separability, ending 
up somewhere close to that border. Finally we used the fact 
that at the middle of collapse intervals the qubits are effectively 
separable from the resonators and the
initial entanglement between the qubits is completely stored 
between the resonators.

We thank Luiz Davidovich for bringing reference \cite{PhysRevA.44.5913} to our attention. We acknowledge financial support from NSF PHY-1203931 

\section{Appendix A}
In this section we calculate the inner products that we used previously to compute the elements of the density matrices. 
\begin{align}
\ipr{\phi_0}{\phi_0}=\sum_{n=0} A_n^2 |r_{n+1}|^2 \\ 
\ipr{\phi_1}{\phi_1}=\sum_{n=0} A_n^2 |t_{n+1}|^2 \\
\ipr{\phi_3}{\phi_3}=\sum_{n=0} A_n^2 |r_{n}|^2 \\ 
\ipr{\phi_2}{\phi_2}=\sum_{n=0} A_n^2 |t_{n}|^2 
\end{align}
 
The next two inner products are 
 \begin{align}
\ipr{\phi_0}{\phi_1}=&\sum_{n=0} A_{n+1}A_n r_{n+2}^* t_{n+1} \\
\ipr{\phi_2}{\phi_3}=&\sum_{n=0} A_{n+1}A_n t_{n+1}^* r_{n} 
\end{align}

and finally we have another four inner products
 \begin{align}
\ipr{\phi_0}{\phi_2}=&\sum_{n=0} A_n A_{n+1} r_{n+1}^* t_{n+1} \\
\ipr{\phi_0}{\phi_3}=&\sum_{n=0}  A_n^2  r_{n+1}^* r_n\\
\ipr{\phi_1}{\phi_2}=&\sum_{n=0} A_n A_{n+2} t_{n+1}^* t_{n+2} \\
\ipr{\phi_1}{\phi_3}=&\sum_{n=0} A_n A_{n+1} t_{n+1}^* r_{n+1}
\end{align}
We can compute the above inner products using appropriate approximations. The first term that we compute is $\ipr{\phi_2}{\phi_2}$.
 \begin{align}
\ipr{\phi_2}{\phi_2}=&\sum_{n=0} A_n^2 |t_{n}|^2 =\sum_{n=0} A_n^2 |\sin(gt\sqrt{n})|^2\\ \nn
=&\frac{1}{2}- \frac{1}{2}\sum_{n=0} A_n^2 \cos(2gt\sqrt{n})
\end{align}
To evaluate the sum we use the approach developed by \citet{PhysRevA.47.4258}. First we rewrite the sum using the Poisson sum formula.
 \begin{align}
\sum_{m=0}^{\infty} f_m=\sum_{\nu=-\infty}^{\infty} \int_{0}^{\infty}dm f(m) e^{2\pi i \nu m}+\frac{f_0}{2}
\end{align}
where $f(m)$ is a continuous version of $f_m$. First we follow the general formalism. In these integrals $A_n^2$ is the Poisson distribution and it only contributes significantly when $m\approx \alpha^2$ and it is the slowly varying factor. Thus we rewrite the sum in $\ipr{\phi_2}{\phi_2}$ as
 \begin{align}
&-\frac{1}{2}\sum_{m=0}^{\infty} A_m^2 \cos(2gt\sqrt{m})\\ \nn
&=-\frac{1}{2} \Re\left(\sum_{m=0}^{\infty}  A_m^2 \exp(-2igt\sqrt{m})    \right)\\ \nn
&=-\frac{1}{2}\Re\left(\sum_{\nu=-\infty}^{\infty} \int_{0}^{\infty}dm A(m)^2  e^{-2igt\sqrt{m}} e^{2\pi i \nu m}+\frac{A_0^2}{2} \right)\\ \nn
&=-\frac{1}{2}\Re\left(\sum_{\nu=-\infty}^{\infty} \int_{0}^{\infty}dm A(m)^2  e^{2iS_{\nu}(m)}+\frac{A_0^2}{2} \right)\\ \nn
&=\sum_{\nu=-\infty}^{\infty}w_{\nu}(t)+\frac{A_0^2}{2}
\end{align}
where $S_{\nu}(m)=\pi\nu m-gt\sqrt{m}$. Now we can use the stationary phase approximation, a variation of saddle point approximation, to approximate these integrals. We approximate $S_{\nu}(m)$ around the point of stationary phase: 
\begin{align}
S_{\nu}(m)=S_{\nu}(m_{\nu})+\frac{1}{2} \frac{\partial^2 S_{\nu}(m_{\nu}) }{\partial m^2}(m-m_{\nu})^2.
\end{align}
where $ \frac{\partial S_{\nu}(m_{\nu}) }{\partial m}=0$. Then we will have 
\begin{align}
w_{\nu}(t)=\frac{-A(m_{\nu})^2}{2}\sqrt{\frac{\pi}{|\frac{\partial^2 S_{\nu}(m_{\nu}) }{\partial m^2}|}}\cos(2S_{\nu}(m_{\nu}) +\eta\frac{\pi}{4})
\end{align}
where $\eta=\text{sgn}(\frac{\partial^2 S_{\nu}(m_{\nu}) }{\partial m^2})$. In the above case $m_{\nu}=\frac{g^2 t^2}{4\pi^2\nu^2}$ and 
\begin{align}
&S_{\nu}(m_{\nu})=-\frac{g^2t^2}{4 \pi \nu}\\
&\frac{\partial^2 S_{\nu}(m_{\nu}) }{\partial m^2}=\frac{2 \pi^2 \nu^3}{g^2 t^2}>0 \\
&w_{\nu}(t)=-\frac{1}{2}A(\frac{g^2 t^2}{4\pi^2\nu^2})^2\left( \frac{gt}{\pi\sqrt{2\nu^3}}  \right)\cos(\frac{g^2t^2}{2 \pi \nu} -\frac{\pi}{4})
\end{align}
The photonic distribution in $A_m^2$ is translated to a series of distributions in time. Since we are interested in positive time, the $\nu=0$ gives the collapse and $\nu>0$ lead to different revivals. Note that for simplicity we can approximate $A_m^2$ with a Gaussian distribution. This is an excellent approximation for moderately large values of $\alpha\ge 10$ that we are interested in here. If we use the above approximation we can conclude that 
 \begin{align}
\ipr{\phi_3}{\phi_3}=\sum_{n=0} A_n^2 |r_{n}|^2 =\frac{1+x}{2} \\ \nn
\ipr{\phi_2}{\phi_2}=\sum_{n=0} A_n^2 |t_{n}|^2 = \frac{1-x}{2}
\end{align}
where $x=\sum_{n=0} A_n^2 \cos(2gt\sqrt{n})$. We also approximate $A_n A_{n+1}=A_n^2$. With this approximation. 
\begin{align}\nn
\ipr{\phi_3}{\phi_3}=\ipr{\phi_0}{\phi_0}\\
\ipr{\phi_2}{\phi_2}=\ipr{\phi_1}{\phi_1}
\end{align}
If we define $y=\sum_{n=0} A_n^2 \sin(2gt\sqrt{n})$ we can estimate 
\begin{align}
\ipr{\phi_0}{\phi_2}=&\sum_{n=0} A_n A_{n+1} r_{n+1}^* t_{n+1}\\ \nn
\simeq& -i\sum_{n=0} A_n^{2} \cos(gt\sqrt{n})\sin(gt\sqrt{n})=\frac{-iy}{2} \\
\ipr{\phi_1}{\phi_3}\simeq& \ipr{\phi_{2}}{\phi_{0}} = \frac{iy}{2}
\end{align}
Now we have only four inner products to calculate and they are all related to each other.
The simplest is 
\begin{align}
\ipr{\phi_0}{\phi_3}=&\sum_{n=0}  A_n^2  r_{n+1}^* r_n\\ \nn 
=&e^{i\omega t} \sum_{n=0}  A_n^2 \cos(gt\sqrt{n+1}) \cos(gt\sqrt{n}),
\end{align}
and we can rewrite this quantity as 
\begin{align}
\ipr{\phi_0}{\phi_3}=&\frac{e^{i\omega t}}{2} \sum_{n=0}  A_n^2 \cos(gt\sqrt{n+1}+gt\sqrt{n})\\ \nn
+&\frac{e^{i\omega t}}{2} \sum_{n=0}  A_n^2\cos(gt\sqrt{n+1}-gt\sqrt{n}) \\ \nn
\simeq& \frac{e^{i\omega t}}{2} \sum_{n=0}  A_n^2 \cos(2gt\sqrt{n}+\frac{gt}{2\sqrt{n}})\\ \nn
+&\frac{e^{i\omega t}}{2} \sum_{n=0}  A_n^2 \cos(\frac{gt}{2\sqrt{n}}).
\end{align}
 Similarly we can get
 \begin{align}
& \ipr{\phi_1}{\phi_2}=\sum_{n=0} A_n A_{n+2} t_{n+1}^* t_{n+2}\\ \nn 
 \simeq & e^{-i\omega t} \sum_{n=0} A_n^{ 2}  \sin(gt\sqrt{n+1}) \sin(gt\sqrt{n})\\ \nn
 \simeq & \frac{e^{-i\omega t}}{2} \sum_{n=0} A_n^{ 2} [\cos(\frac{gt}{2\sqrt{n}})- \cos(2gt\sqrt{n}+\frac{gt}{2\sqrt{n}})]
\end{align}
and
 \begin{align}\nn
&\ipr{\phi_0}{\phi_1}=\sum_{n=0} A_{n+1}A_n r_{n+2}^* t_{n+1} \\ \nn 
\simeq& -i e^{i\omega t} \sum_{n=0} A_n^{2} \cos(gt\sqrt{n+1})\sin(gt\sqrt{n})\\ \nn
=& -i \frac{e^{i\omega t}}{2} \sum_{n=0} A_n^{2}[-\sin(\frac{gt}{2\sqrt{n}})+\sin(2gt\sqrt{n}+\frac{gt}{2\sqrt{n}})],\\
&\ipr{\phi_2}{\phi_3}=\sum_{n=0} A_{n+1}A_n t_{n+1}^* r_{n}\\ \nn
 \simeq& i e^{i\omega t} \sum_{n=0} A_n^{2} \sin(gt\sqrt{n+1})\cos(gt\sqrt{n})\\ \nn
=&\frac{i e^{i\omega t}}{2} \sum_{n=0} A_n^{2} [\sin(\frac{gt}{2\sqrt{n}})+\sin(2gt\sqrt{n}+\frac{gt}{2\sqrt{n}})].
\end{align}
Two of the above summations lead to evaluations of $x,y$ that we already explained how to evaluate. The other two summations are evaluated in \cite{yonac2010}. In the limit of $\alpha^2\gg 1$ we have
\begin{align}
I=\sum_{n=0} A_{n}^{2} \exp(\frac{igt}{2\sqrt{n}})\simeq\exp(-\frac{g^{2}t^{2}}{32\alpha^{4}}) e^{\frac{igt}{2\alpha}}=I_{1}+iI_{2}
\end{align}

\section{Appendix B}

In this section we focus on the entanglement properties of a subclass of GHZ-diagonal states. The GHZ-diagonal states are themselves a subset of X-states. In our notation they are X-states for which $a_i=b_i$. Such matrices can be written as a convex sum of GHZ states, hence their name. The entanglement properties of GHZ-diagonal states has already been the subject of a few previous investigations 
\cite{,PhysRevA.61.042314,1367-2630-12-5-053002,PhysRevA.83.020303,guhne2011,PhysRevA.88.062331}.

For the purpose of the current manuscript we can impose an extra condition of $z_i=0$ for $i>1$. The states that we are interested in are X-matrices of the form given below.
\begin{align}
\hat{\rho}_x=\left(  \begin{array}{cccccccc}
    a_{1} & ¥ & ¥ & ¥ & ¥ & ¥ & ¥ & z_{1} \\ 
    ¥ & b_{2} & ¥  & ¥ & ¥ & ¥ & 0 & ¥ \\ 
    ¥ & ¥ & \ddots & ¥ & ¥ & \iddots & ¥ & ¥ \\ 
    ¥ & ¥ & ¥ & b_{d} & 0 & ¥ & ¥ & ¥ \\ 
    ¥ & ¥ & ¥ & 0 & b_{d} & ¥ & ¥ & ¥ \\ 
    ¥ & ¥ & \iddots & ¥ & ¥ & \ddots & ¥ & ¥ \\ 
    ¥ & 0 & ¥ & ¥ & ¥ & ¥ & b_{2} & ¥ \\ 
    z_{1}^{*} & ¥ & ¥ & ¥ & ¥ & ¥ & ¥ & a_{1} \\ 
  \end{array}
\right).\label{GHZdiagonal}
\end{align}

The conditions for biseparability and full separability of all partitioning of such states are known \cite{PhysRevA.61.042314}. For these 
states to be fully separable it is necessary and sufficient that 
\begin{align}
|z_1|- \min\{b_i\}\le0
\end{align}
Below we show that the quantity $S$ also has a simple geometrical interpretation and it can be used as a measure of weak inseparability. We 
submit that $S\ge0$ is the distance from the closest fully separable state. The distance metric we are using here is the trace distance that
we have used previously to find a measure of all-party entanglement \cite{PhysRevA.88.062331}. The trace distance of two matrices is 
given by
\begin{align}
D(\rho,\tau)=\frac{1}{2}\text{Tr}(|\rho-\tau|),
\end{align}
where $|A|=\sqrt{AA^{\dagger}}$. The measure of weak inseparability that we introduce here is 
\begin{align}
E(\rho)=\min_{\tau\in \mathcal{FS}}2D(\rho,\tau)
\end{align}
where $\mathcal{FS}$ denotes the set of fully separable states, which is a convex set. Identically to the proof we have given in \cite{PhysRevA.88.062331}
we can show that $E(\rho)$ is convex, non-increasing under LOCC and invariant under local unitary transformations. From now on, unless otherwise said, we use separable in stead of fully separable. Now we prove the main result of this section that is to find the closest fully separable state to $\hat{\rho}_x$. It is shown in \cite{ PhysRevA.61.042314} that any density matrix can be depolarized to a state of the form $\hat{\rho}_x$ using an LOCC map. This and the contractive property of the trace distance guarantee that the closest separable state to $\hat{\rho}_x$ has the same form as $\hat{\rho}_x$. In the following we assume that $z_1=|z_{1}|$ without loss of generality since this can always be accomplished using a local unitary transformation without changing any other element of the matrix, and the distance is invariant under local unitary transformaitons. We assume $z_1=\min\{b_i\}+\epsilon>\min\{b_i\}=c$. We will prove that the closest separable state to $\hat{\rho}_x$ has identical elements except for $z_1$ replaced with with $c$, i.e. $E(\hat{\rho}_x)=\epsilon$. 

To prove let us assume the contrary. There exists a separable state $\hat{\Sigma}$ that is closer to $\hat{\rho}_x$ than $\epsilon$. We parameterize this state as $\hat{\Sigma}=\hat{\rho}_x+\hat{\Delta}$:

\begin{align}
\hat{\Delta}=\left(  \begin{array}{cccccccc}
    \delta_{1} & ¥ & ¥ & ¥ & ¥ & ¥ & ¥ & \nu-\epsilon \\ 
    ¥ & \delta_{2} & ¥  & ¥ & ¥ & ¥ & 0 & ¥ \\ 
    ¥ & ¥ & \ddots & ¥ & ¥ & \iddots & ¥ & ¥ \\ 
    ¥ & ¥ & ¥ & \delta_{d} & 0 & ¥ & ¥ & ¥ \\ 
    ¥ & ¥ & ¥ & 0 & \delta_{d} & ¥ & ¥ & ¥ \\ 
    ¥ & ¥ & \iddots & ¥ & ¥ & \ddots & ¥ & ¥ \\ 
    ¥ & 0 & ¥ & ¥ & ¥ & ¥ & \delta_{2} & ¥ \\ 
    \nu^{*}-\epsilon & ¥ & ¥ & ¥ & ¥ & ¥ & ¥ & \delta_{1} \\ 
  \end{array}
\right).
\end{align}

The $X$ form of the difference matrix implies that $D(\hat{\rho}_x,\Sigma)=\frac{|D_1|}{2}+\Sigma_{i>0} |\delta_i| $.\\
\begin{align}
D_1=\left(\begin{array}{cc}
\delta_{1} & \nu_{1}-\epsilon     \\
 \nu_{1}^{*} - \epsilon & \delta_{1}   \\
\end{array}
\right)
\end{align}
The contribution from $D_1$ leads to 
\begin{align}
\frac{|D_1|}{2}\ge|\epsilon-\nu|\ge|\epsilon-\Re(\nu)|.
\end{align}
The above inequality implies that $\Re(\nu)>0$ since we assumes that $D(\hat{\rho}_x,\Sigma)<\epsilon$. Since $\Sigma$ 
is a separable state we should have that for all $i$'s
\begin{align}
b_i+\delta_i\ge |c+\nu|\ge c+\Re(\nu).
\end{align}
Let $i=j$ be the index for which $b_j=c$, then $\delta_i \ge \Re(\nu)$. Thus we have $D(\hat{\rho}_x,\Sigma)\ge |D_1|+\delta_i\ge \epsilon$. This implies that there is no separable state closest to $\hat{\rho}_x$ than $z_1-c$. The state with $\delta_i=0$ and $\nu=0$ saturates this inequality and thus $E(\rho)=\min\{|z_1|-c,0\}$. Note that in the manuscript we choose the normalization $S=2E(\hat{\rho}_x)$ so that the distance matches the value of the all-party concurrence for $\hat{\rho}_x$.

\bibliographystyle{apsrev4-1}
\bibliography{mybib}

\end{document}